\documentclass[runningheads]{llncs}
\usepackage{url}
\usepackage[pdftex]{graphicx}
\usepackage{todonotes}
\usepackage{float}
\usepackage{booktabs}
\usepackage[pdftex,bookmarks=true,plainpages=false,pdfpagelabels=true]{hyperref}
\usepackage{enumerate}
\usepackage{paralist}
\usepackage{array}
\usepackage{longtable}
\usepackage{listings}
\usepackage[utf8]{inputenc}
\usepackage[capitalize, noabbrev]{cleveref}
\usepackage{tabularx}
\usepackage{multirow}
\usepackage{bigstrut}
\usepackage{verbatim}
\usepackage{geometry}

\newcolumntype{L}[1]{>{\raggedright\let\newline\\\arraybackslash\hspace{0pt}}p{#1}}
\newcolumntype{C}[1]{>{\centering\let\newline\\\arraybackslash\hspace{0pt}}p{#1}}
\newcolumntype{R}[1]{>{\raggedleft\let\newline\\\arraybackslash\hspace{0pt}}p{#1}}

\begin{document}
\title{Integration of Security Standards in DevOps Pipelines: An Industry Case Study (Author\'s Copy)
}
%
%
\author{Fabiola Moy\'on\inst{1}\orcidID{0000-0003-0535-1371}
\and \newline {Rafael Soares} \inst{2} \and
Maria Pinto-Albuquerque\inst{2}\orcidID{0000-0002-2725-7629}\and \newline
Daniel Mendez \inst{3}\orcidID{0000-0003-0619-6027}  \and   
Kristian Beckers\inst{4} } 
%
\authorrunning{Author\'s Copy}
%
\institute{Siemens CT and Technical University of Munich, Germany \email{fabiola.moyon@siemens.com}
\and
Instituto Universitário de Lisboa (ISCTE-IUL), ISTAR-IUL, Lisboa, Portugal \newline  
 \email{rafael\_soares@iscte-iul.pt, maria.albuquerque@iscte-iul.pt}
\and
Blekinge Institute of Technology, Sweden, and fortiss GmbH, Germany \newline \email{daniel.mendez@bth.se}
 \and
Social Engineering Academy, Munich, Germany \email{kristian.beckers@social-engineering.academy}
}
\maketitle              
\begin{abstract}
In the last decade, companies adopted DevOps as a fast path to deliver software products according to customer expectations, with well aligned teams and in continuous cycles. As a basic practice, DevOps relies on pipelines that simulate factory swim-lanes. The more automation in the pipeline, the shorter a lead time is supposed to be. However, applying DevOps is challenging, particularly for industrial control systems (ICS) that support critical infrastructures and that must obey to rigorous requirements from security regulations and standards. Current research on security compliant DevOps presents open gaps for this particular domain and in general for systematic application of security standards.
In this paper, we present a systematic approach to integrate standard-based security activities into DevOps pipelines and highlight their automation potential. Our intention is to share our experiences and help practitioners to overcome the trade-off between adding security activities into the development process and keeping a short lead time. We conducted an evaluation of our approach at a large industrial company considering the IEC 62443-4-1 security standard that regulates ICS. The results strengthen our confidence in the usefulness of our approach and artefacts, and in that they can support practitioners to achieve security compliance while preserving agility including short lead times.

\keywords{Secure Software Engineering \and Security Standards \and Agile Software Engineering \and DevOps pipeline \and DevSecOps \and Industrial Control Systems}
\end{abstract}
\section{Introduction}\label{sec:intro}

Agile methodologies aim to deliver software products that satisfy customer needs while enabling collaboration among stakeholders \cite{beck2001manifesto}. 
Lean techniques apply manufacturing flows to deliver software products with waste reduction, increased visibility of the manufacturing pipeline, and better team collaboration \cite{poppendieck:2003:lean}
DevOps relies on both agile and lean practices to break the barriers between development (Dev) and operation (Ops) teams \cite{Jabbari:2016}. This extends the benefits beyond delivery to the operation of software products. By applying DevOps, organizations attempt to deliver and operate software products according to customer expectations, with well aligned teams and with focus on continuous improvement of the process flows. The indicator of improvement is reducing the time-frame to transform a customer need into a usable software functionality at the production environment -- so-called lead time \cite{kim:2016:devops}.

To shorten the lead time, DevOps relies on automation practices \cite{leite:2019:survey},
where Continuous Integration/Continuous Delivery (CI/CD)\label{label_cicd} pipelines are essential. The term \emph{pipeline} refers to so-called factory swim-lanes and describes a systematic alignment of processes and tools to release software products in a seamless manner, often characterised with the metaphor of "pushing a button". This "button" triggers a set of automated checks and tests aimed at software quality assurance \cite{humble2010continuous}.

While DevOps was originally conceptualized for IT companies, industrial companies have started as well to embrace DevOps practices\cite{allspaw:2009:ten,dora_report}. However, there are yet several challenges to overcome before DevOps can be largely applied in highly regulated domains with high demand for quality attributes, in particular security \cite{yasar:2017:implementing}. 

Especially the domain of industrial control systems (ICS)\label{label_ics}, where software products are vital to support critical infrastructure, is characterised by the need of compliance with security standards. The ICS domain is regulated by the security standards family IEC-62443, where the IEC-62443-4-1 standard (the 4-1 standard)\label{label_41} states process requirements for secure product development life-cycle \cite{iec4_1}. 

Nowadays, the software engineering (SE) field lacks methods to demonstrate compliance with security standards when applying DevOps \cite{yasar:2017:implementing}. There are initial contributions on security in agile methods, as well as on DevOps (e.g. \cite{michener:2016:mitigating,jaatun:2017:devops,yasar:2016:where,rahman:2016:software});
however, publications which address the compliance of security are very scarce and do not yet cover completely the ICS domain \cite{laukkarinen:2018:regulated,morales:2020:guide}.

In addition, a challenge for the SE field is the trade-off between adding security activities during development and keeping a short lead time \cite{kim:2018:phoenix}.

To achieve security compliance for DevOps, we argue that we first need to understand how to systematically integrate security standard requirements into DevOps pipelines. In this paper, we address this problem for the representative standard in the ICS domain: the IEC 62443-4-1 standard for secure product development. Through a case study, we analyze its application in a large industrial company.  

 Integration is possible since both the security standard and the DevOps pipeline are based on activity flows. The 4-1 standard requirements are a set of processes to ensure security in the product development lifeycle. The DevOps pipeline is a process chain to deliver software products \cite{humble2010continuous}. Hence, we mapped the 4-1 process requirements into the applicable DevOps pipeline stages, namely: concept, code, build, test, release, deploy, operate and monitor \cite{Jabbari:2016}. 

In addition, given that DevOps relies on automation to keep short lead times, we determine to what extent the 4-1 standard requirements can be automated. Ideally, the more \textit{automatable} the standard security requirement, the less impact on the lead time. 

In summary, to improve the product development process in security regulated environments, this work presents two contributions:
\begin{enumerate}
    \item an approach to systematically describe security compliant DevOps pipelines, together with a first instance for the IEC 62443-4-1 standard, and
    \item a description of automation capabilities of the 4-1 standard security requirements, together with details of available security tools, if one exists to date. 
\end{enumerate}
 
Our contributions may support companies driving DevOps for ICS to satisfy not only customer needs but also regulatory demands, without losing the benefits of agile and DevOps.  With a description of \textit{non-automatable} standard requirements, we raise awareness of where in the product life-cycle to emphasize collaboration between security and DevOps teams. Also, by describing automation capabilities, we provide paths to implement security with less impact on the lead time. Moreover, for the research field of security compliance, our work reduces the gaps with relation to security compliant DevOps.

To evaluate the applicability and usefulness of our contributions, we conducted a qualitative study consisting of interviews with expert practitioners in a large industrial company. Results show that the integration approach may be applied to describe DevOps compliance with other security standards or in other regulated domains, e.g. finance, telecommunications. Moreover, the study revealed that the 4-1 standard automation capabilities motivate practitioners to implement compliance programs based on automation. Nowadays, this work's artefacts are applied in the company to introduce Dev and Ops teams to security compliance with the 4-1 standard and to perform DevSecOps assessments.

The rest of this paper is structured as follows. In Section 2, we discuss fundamentals and related work. Section 3 presents the steps we propose to integrate security standards into DevOps pipelines, through systematic identification of activities and automation capabilities, while pointing out at which stages of the pipeline each security requirement fits. Section 4 presents how the integration approach delivers, for the 4-1 standard, a description of automation capabilities as well as a specification of Security Standard Compliant (S\textsuperscript{2}C) DevOps Pipeline Specification. Section 5 reports on the evaluation. In Sections 6, we discuss the main findings, impact and limitations of our work. Finally in section 7, we summarize current and further work. \par

\section{Fundamentals and Related Work}\label{sec:fundamentals}
In this section, we describe the DevOps concepts including the automation practice. Later, we present a summary of the relevant work in the field of security compliant DevOps.

\subsection{DevOps and Pipelines}

DevOps combines working philosophies and practices to remove the barrier between the development (Dev) and operation (Ops) teams. Key elements include collaboration, automation, measurement, and monitoring \cite{lwakatare2015dimensions}. An essential aspect for automation is the concept of the Continuous Integration/Continuous Delivery (CI/CD) pipeline. This pipeline describes the alignment of processes and tools to automate steps in the Software Development Life Cycle (SDLC) process, which includes initiating code builds, running acceptance tests, and deploying to staging and production environments. Pipelines produce artefacts that may serve as evidence for compliance \cite{humble2010continuous}.

Debates on DevOps concepts, definitions, and practices are active with yet no clear consensus \cite{leite:2019:survey}. In the contribution at hands, we concentrate on the so-called \textit{automation practice} \cite{Jabbari:2016} and we recognize pipelines as the concrete manifestation of this practice in industrial environments. When using the term DevOps pipeline, we refer to the sequence of processes that transform needs of a customer into valuable product increments and deployed them to production site. 

For this work, we chose as main reference the original DevOps Pipeline (the well-known infinite symbol), as it represents the sequence from customer needs to deployed functionalities. For specific details, we analyzed also the pipeline proposals from: Bird, Humbley and Gartner \cite{bird2016devopssec,humble2010continuous,gartner_report}.

\subsection{DevSecOps and Security Standards}

The term "DevSecOps" has emerged as organisations, that started to apply DevOps techniques, were concerned about security aspects. It refers to the incorporation of security practices in a DevOps environment through the collaboration between development, operation, and security teams \cite{mohan2016secdevops}. Reports correlated security automation with DevOps success and recommended the integration of security earlier in the development life cycle, moving from operational to development stages \cite{dora_report,sonatype_report}. To achieve this, CI/CD pipelines were adapted to include security practices \cite{bird2016devopssec}.

In general, publications, that refer to DevOps and regulations-based security, are scarce. Authors explore security in regulated environments \cite{yasar:2017:implementing,michener:2016:mitigating} and where to introduce security activities in DevOps \cite{yasar:2016:where,jaatun:2017:devops}. However DevSecOps compliant with a particular standard or domain is still missing. In our experience in industrial environments, a major gap for security compliance is identifying which DevOps artefacts can serve as compliance evidence. 

The ICS domain is regulated by the IEC 62443 standard family, whose sibling the IEC 62443-4-1 (the 4-1) provides process requirements (activities, artefacts, and flows) to achieve a secure product development life-cycle \cite{iec4_1}. It contains eight practices, which are: \textbf{security management} (SM), to ensure that security activities are adequately executed through the product's life cycle; \textbf{specification of security requirements} (SR), to accurately elicit product's security capabilities; \textbf{secure design} (SD), to ensure that security is involved from overall architecture to individual components; \textbf{secure implementation} (SI), to ensure applicability of secure coding and implementation practices; \textbf{security verification and validation testing} (SVV), to ensure that security design was implemented; \textbf{management of security-related issues} (DM), to handle product's security-related issues; \textbf{security update management} (SUM), to ensure timely delivery of security updates; and finally \textbf{security guidelines} (SG) to provide sufficient documentation for secure product deployment. 


This work extends the field by specifying which practices of the 4-1 standard apply for each DevOps phase. Although previous work refers to security involvement into pipelines \cite{bird2016devopssec}, our contribution fills the gap of applying security from the point of view of security standards: involving more than technology, the people and process aspects. In addition, at a granular level, we determine which security standard activities can be automated in a pipeline.  

\section{Integration of Security standards into DevOps Pipelines}\label{sec:integration}
In industrial environments adopting DevOps, we aim to improve the product development process by achieving security compliance with less impact on the leadtime. Therefore it is our intention to determine to what extent security standard requirements can be automated and how a DevOps pipeline will look like when orchestrating such requirements.

\begin{figure*}[!htb]
 \centering
  \includegraphics[width=0.85\linewidth]{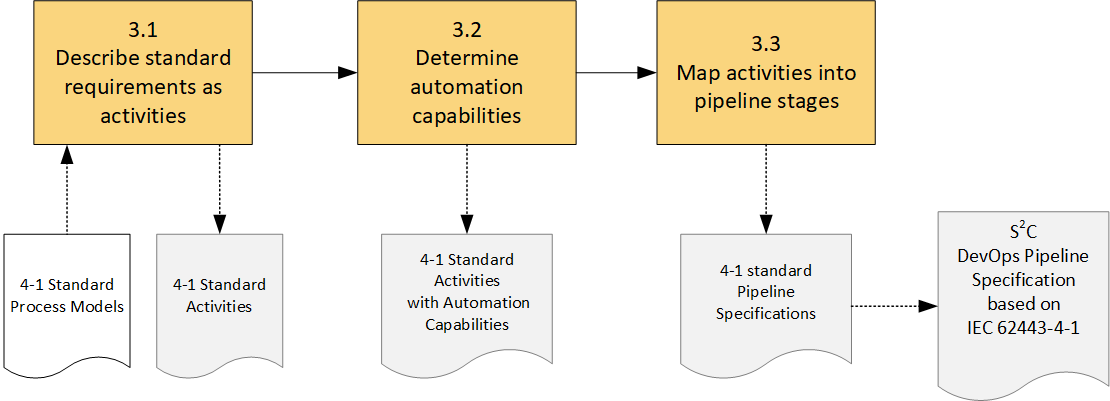}
  \caption{Approach to integrate security standards into DevOps Pipelines. 
  For illustration purposes, artefacts are shown as particular instances of the 4-1 standard. Artefact in white was adopted from previous work \cite{Moyon:2018}. The final output of this approach is the Security Standard Compliant (S\textsuperscript{2}C) DevOps Pipeline presented in \cref{sec:concept}}
  \vspace{-10pt}
  \label{fig:03_integration}
\end{figure*}

Security compliance requirements are stated in security standards and to integrate them into DevOps, we propose a systematic approach that consists of three steps (see \cref{fig:03_integration}). First, we list the standard activities in a precise way. Second, for each standard activity, we determine the automation capabilities and finally, we map activities into the DevOps Pipeline stages. The approach is applied for the 4-1 security standard for secure product development in ICS. The artefacts presented in this paper are instances for this particular standard; however, the structure can serve as template for the analysis of other standards describing secure development life-cycles and applicable in other domains like the ISO 27034 for secure software development. A complete set of artefacts is part of our contribution and can be accessed in our online material at \url{https://doi.org/10.6084/m9.figshare.11294534}.

In the following sub-sections, we describe the steps in detail and present the artefacts. 

\subsection{Describe standard requirements as activities} This step focuses on a detailed analysis of the standard and its requirements. As result, we obtain a precise description of the standard requirements to the level of activities with inputs and outputs. Each requirement may contain several \textit{activities} to be orchestrated in a pipeline. Inputs and outputs serve to build up the orchestration flow, meaning: the output of one activity is the input of the next. 

For this case study, we based this analysis on existent process models of the 4-1 standard~\cite{Moyon:2018}. Such models represent the 4-1 standard requirements with the Business Process Model and Notation (BPMN).  
From the 4-1 process models, we extracted tasks, events, and gateways of the standard requirements. All of them where considered as \textit{activities} to be orchestrated. Further, we also made explicit the input and output based on the artefacts, also depicted in the 4-1 process models. This analysis resulted into 160 activities.  

\begin{figure*}[!h]
 \centering
  \includegraphics[width=0.8\textwidth,keepaspectratio]{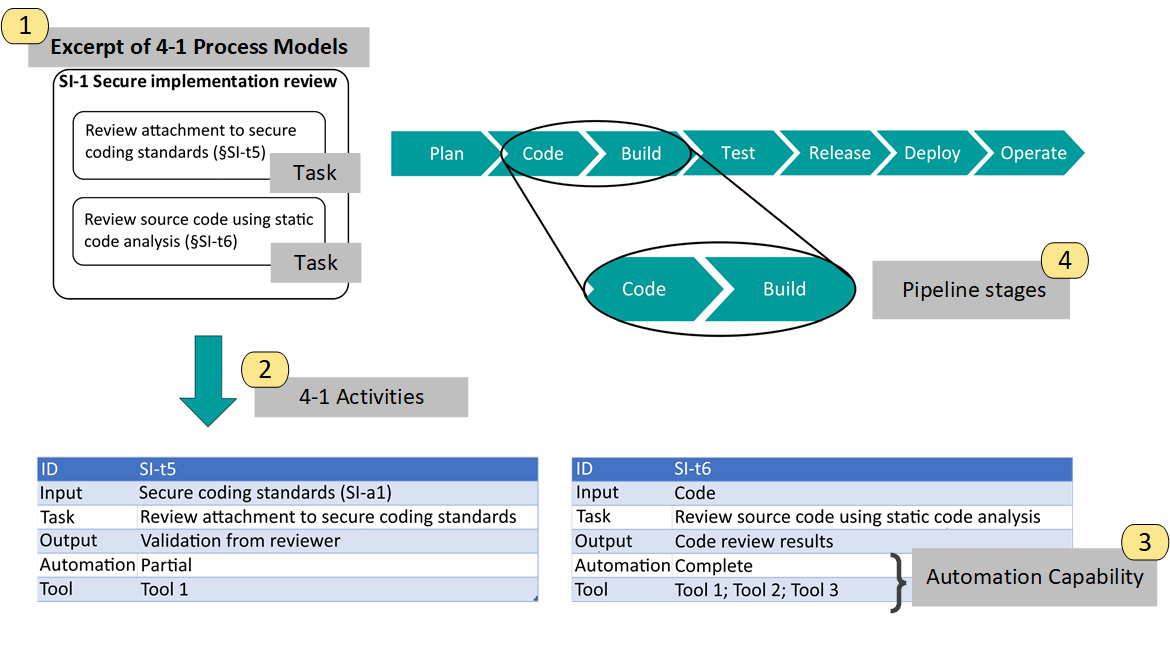}
  \caption{Description on how IEC 62443-4-1 requirements extracted from process models (1) can be refined as activities (2) with input, output and automation capabilities (3). Finally, the activities are mapped into pipeline stages (4)}
    \vspace{-10pt}
  \label{fig:excerpt_specification}
\end{figure*}

Figure \ref{fig:excerpt_specification} (steps 1 and 2) presents an example of how activities are extracted from a requirement of the 4-1 standard. The example shows an excerpt of how the 4-1 process models depict the requirement \textit{SI-1 Secure implementation review} belonging to the practice Secure Implementation.  

\subsection{Determine automation capabilities for the standard}  This step analyzes if the standard activities can be automated. The term \emph{automation capability} is based in two criteria: 

\subsubsection{Automation Level} Describes to what extent the security standard activities can be automated. The following are the \emph{automation categories}:
    \begin{itemize}
        \item[-] Human Task: Automation is not possible. A human must perform the activity.
        \item[-] Transparency: A human can perform the activity based on tool results e.g. visualization.
        \item[-] Partial Automation: Parts of the activity can be automated but required manual input to be completed.
        \item[-] Tool Possible: Activity can be automated, but no tool was identified 
		\item[-] Complete: Activity can be completely performed by a tool and tools are available
    \end{itemize}
\textbf{Tool support} For activities that can be automated, we searched available tools in several sources and investigate if the tools fit the automation level \cite{hsu2018hands,sansposter,shahin2017continuous}. 
Finally, a tool list is compiled including first open source tools as well as those that can be integrated into Continuous Integration tools (e.g.,\textit{Jenkins}, \textit{Gitlab CI}). 
    
The automation level and tool support are dependent criteria, e.g. an activity may seem to be fully automated, however there may not be a tool that fulfills the activity completely.
Figure \ref{fig:excerpt_specification} (step 3) shows an example of how automation capabilities are described for the activities extracted after step 3.1. These activities refer two tasks that implement the requirement SI-1 Secure implementation review. The 4-1 process models notate them as SI-t5 and SI-t6. 

\subsection{Map activities into pipeline stages }  
In this step, we identify in which stages of the DevOps pipeline the security standard activities should take place. To this aim, we find characteristics in common in both DevOps pipeline and the 4-1 standard. Afterwards, we determine for each activity of the 4-1 standard the corresponding stage(s) of the DevOps pipeline.  Security standard activities are included as early as possible in the pipeline, prioritizing stages where the product is not yet in the production environment. \cref{fig:excerpt_specification} (step 4) shows an example of the mapping of 4-1 activities into pipeline stages. 
To have an overview of every 4-1 practice, the individual activities with their automation capabilities are aggregated into per 4-1 practice. These models are called \textit{Pipeline Specifications of the Standard Practice} and are available in the online material. 

Finally, this step aggregates the individual activities in a high-level overview, a Security Standard Compliant (S\textsuperscript{2}C) DevOps Pipeline. In this section, we presented an example for the SI practice of the 4-1 standard. In the following section, we point out key points related to the rest of the standard practices.


\section{DevOps compliant with the 4-1 Security Standard}\label{sec:concept}
Applying the integration steps for the 4-1 security standard resulted into: a description of automation capabilities for the 4-1 standard practices and a instance of a (S\textsuperscript{2}C) DevOps Pipeline for this specific standard. In this section, we describe them.%
\subsection{Automation Capabilities of the 4-1 Standard}

Ideally, to support the DevOps aim of reducing the lead time, security activities require to be completely automated. Summarizing the 4-1 security activities per automation level, this occurs for 31\% of the 4-1 security activities. It means that a tool is available and can completely implement the security activity when orchestrated into a pipeline. The opposite occurs for 38\% of the standard activities, which cannot be automated at all and where a human must do the task. Hence, these security activities require comprehensive collaboration among stakeholders, either including a security expert into the iterations or providing security knowledge to DevOps team members.

\begin{figure}[!hb]
 \centering
  \includegraphics[width=0.85\linewidth,keepaspectratio]{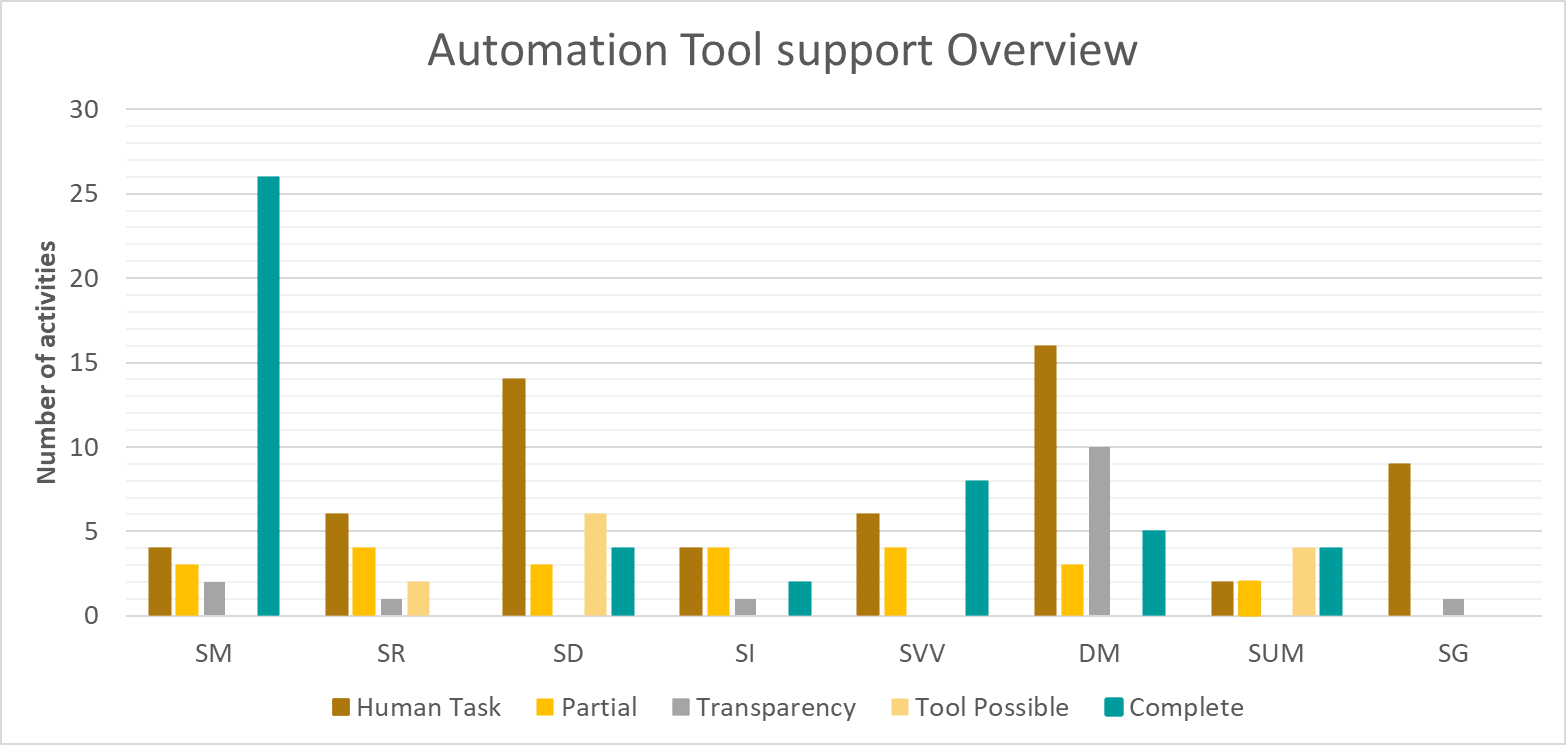}
  \caption{Summary of Automation capabilities per practice of the IEC 62443-4-1. For the whole standard, this corresponds to: Human Task 38\%; Transparency 9\%; Partial Automation 14\%; Tool Possible 8\%; Complete Automation 31\%   
  }
  \label{fig:automation_stats_practice}
\end{figure}

Some 4-1 activities (9\%) can be supported from tools to achieve transparency, meaning that a human can take decisions based on automation e.g. dashboard tools that aggregate security findings. Other 14\% of security activities can be partially implemented by an existent security tool. Finally, 8\% of standard security activities could be automated but, to the best of our understanding, there is not yet documentation in our sources that such a tool is implemented and used, e.g. tool for automatic aggregation of security findings and prioritization into the backlog based on threat models.  
\cref{fig:automation_stats_practice} shows the automation capabilities of the 4-1 activities grouped per practice. The 4-1 standard practices with high automation capabilities are: security management (SM), secure verification and validations testing (SVV), and security update management (SUM). For SM such automation level may not be clear at first sight, but the 4-1 standard includes in this practice  security activities for environment verification, encryption/key management and vulnerability checking of third-party components. Several security tools implement these use cases. The SVV practice reflects a common use case, security testing activities like vulnerability checking or port scanning are well automated and supported by tools. For the SUM practice, tools fully automate security activities for updates delivery and installation.

In contrast, we observe practices where no automation is possible: security requirements (SR) and security guidelines (SG). For SR, most security activities rely on human effort (see human and partial automation in \cref{fig:automation_stats_practice}). DevOps Teams should find suitable requirements engineering techniques to support DevOps objectives. In our experience, compliance of security requirements practices is challenging for industrial environments applying agile and DevOps practices. 
For SG, security activities are manual since they refer to documentation of secure configuration guidelines
.  In addition, we observe some practices largely depending on human tasks, such as secure design (SD) and management of security-related issues (DM). In SD, humans have to generate security architecture diagrams or define measures. Tools can help for drawing or to auto-generate architecture models but humans need to approve them. In DM, security tools aggregate issues but experts decide what to do with findings.

\subsection{Security Standard Compliant DevOps Pipeline}
Besides automation capabilities, we contribute with a specification of how security standard activities should be involved into DevOps stages. \cref{fig:s2c_pipeline} shows the (S\textsuperscript{2}C) DevOps Pipeline for the 4-1 standard and a brief description of key points is the following:

\begin{figure*}[h]
	\centering
	\includegraphics[width=\textwidth]{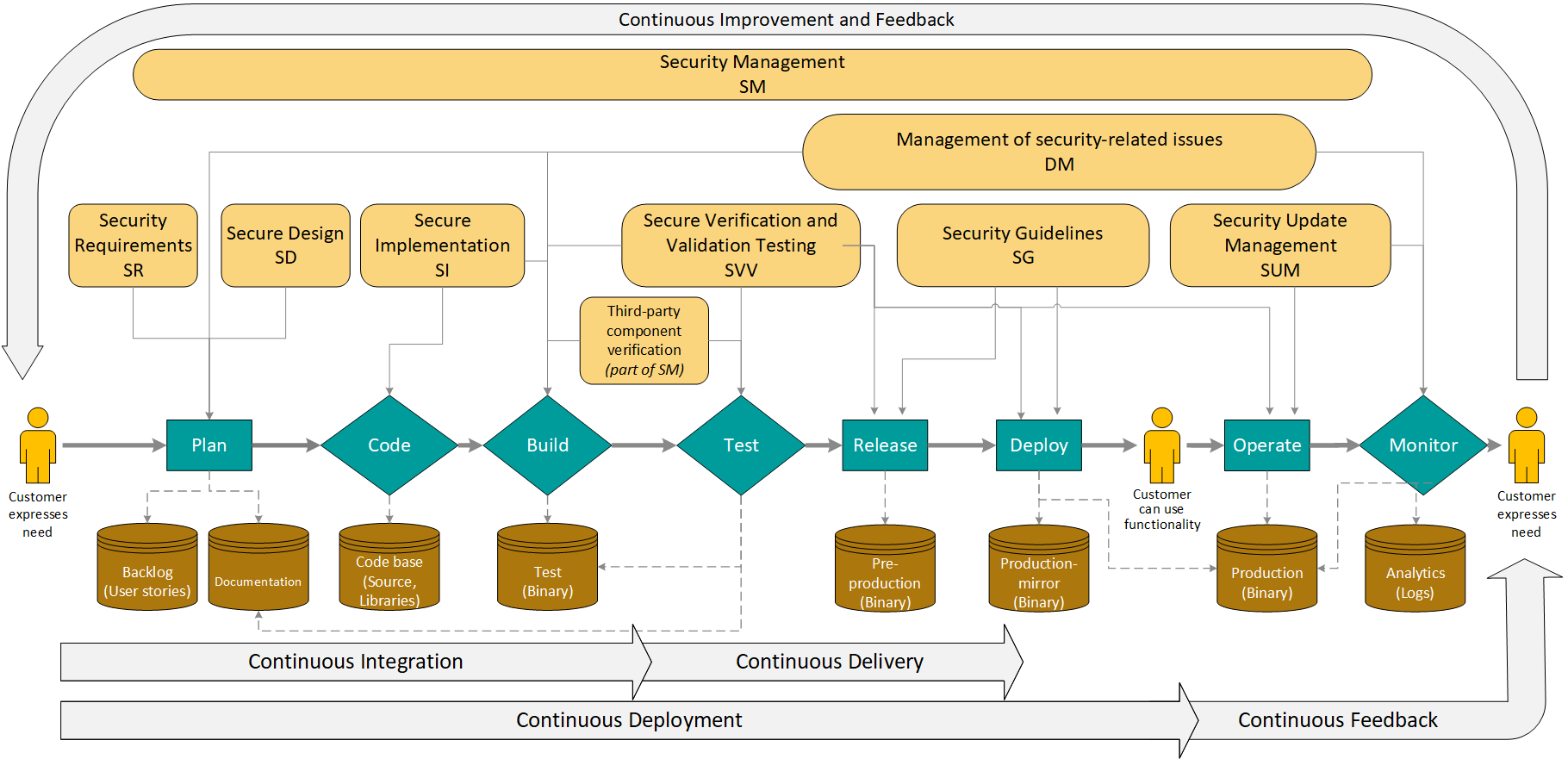}
	\small \caption{Security Standard Compliant DevOps Pipeline for the IEC 62443-4-1 Standard: Diagram shows the 4-1 standard practices (in yellow), the DevOps stages (in green). Solid vertical arrows depict in which DevOps stage a 4-1 practice security activity can take place. Standard security activities impact several repositories (in brown) like backlog, code based and test, pre-production and production environments. In addition, security standard demands an explicit repository for documentation and logs maintenance. Continuous practices (gray arrows) describe the flows to which the security activities also apply.}
	\label{fig:s2c_pipeline}
\end{figure*}

\vspace{-5pt}
\subsubsection{Plan:} Based on the customer need, during this stage, requirements are represented as user stories. Security requirements (SR) should be included in the elicitation flow. The activities of the SR practice are rather manual and no automation is possible (see \cref{fig:automation_stats_practice}). 

The 4-1 Standard demands to perform \textit{threat modeling} and describe the \textit{product security context}. The latter refers to describing required security measures for the environment where the product will be deployed, e.g. networks or operating systems. 

To perform these activities, teams need to involve security specialists who recommend modeling tools. Such tools do not automate the activities but provide transparency and support documentation. Some threat modeling tools provide lists of common threats which consider partial automation as the experts do not need to list the threats. 

A key part for compliance is to ensure that these activities are performed, therefore the backlog management tool should include them. Here, we identified potential for automation. User stories referring to critical components (as based on threat models) could be automatically labeled by the backlog management tool as security relevant. Automatic labeling can be based on natural language analysis. This will facilitate collaboration between DevOps teams and security experts, since security experts are included by default in the process.

In addition, in this stage, the 4-1 Practice Secure Design (SD) demands to define \textit{secure design best practices} and \textit{secure design principles}. These are later transformed into security measures for each \textit{interface} of the product. Although most activities are manual, automation may help to add preset lists of security measures that can be included into the design. Repositories that provide compliance evidence are the Backlog and Documentation repository. 

Finally, the practice Management of Security-related issues (DM) demands that if  \textit{security issues} are detected at any stage of the pipeline, the solution should include updates in requirements and design. Although automation exists to detect, track and manage security issues, the manual effort is high specially for disclosure procedures and issues validation and investigation.
\vspace{-10pt}
\subsubsection{Code:} At this stage, teams code functionalities based on the concept. DevOps establishes that not only  functionalities can be coded, but also infrastructure and environment configurations (the shifting left concept where Operation Teams start working at the Code phase \cite{kim:2016:devops}. At this point, the 4-1 security standard demands secure implementation (SI). Automation is possible for static code analysis. Other standard activities are rather manual like definition of coding standards and review of attachment to coding standards e.g. through peer-review.  

DevOps recommends that functionalities and configurations should be stored in the same repository and integrated into the version control tool. This allows developers to pull not only code but also an environment that is very similar or mirror to production. From the 4-1 standard point of view, this is relevant since it allows to maintain evidence of all security measures including environment hardening. 
In addition, it enables compliance with the standard requirements of development environment protection part of the Security Management practice (SM).

\subsubsection{Build:} During this stage, the teams (not only developers) commit their code. The continuous integration tool triggers different security testing tools. If testing is successful, code is merged and the application is ready for testing. If testing is not successful, commit is not merged and results of testing tools need to be synchronized with the security issues tracking tool to comply with the 4-1 standard DM Practice. 

The 4-1 practice Secure Validation and Verification testing  (SVV) describes four groups of testing: \textit{security requirements testing}, \textit{threat mitigation testing}, \textit{vulnerability testing} and \textit{penetration testing}. During Build, security requirements testing can be done with development frameworks for behaviour-driven development and unit testing. Threat mitigation testing requires manual tasks. Vulnerability testing can be completely automated with specific tools depending on the programming language. 

The 4-1 practice Security Management (SM) contains the requirement: \textit{third party component verification}. Several tools provide support for this activity. Checks avoid the use of vulnerable components as part of the system architecture. This process is known as Continuous Integration
\vspace{-5pt}
\subsubsection{Test:} In the test environment, automated and user acceptance tests can occur. The activities of the 4-1 practice SVV as well as third party component verification apply and are highly automated. To achieve compliance, testing documentation should exist and located in the documentation repository. 
\vspace{-10pt}
\subsubsection{Release and Deploy:} During these phases, the functionalities that implement the customer's need are made available for customers. The release stage refers to Alfa/Beta releases and stored into a pre-production repository and during the Deploy stage to the production environment. Releases allow to gather early feedback about the implemented security measures. 

During these stages, penetration testing is performed with tools that partially automate the activities. Tools run tests, identify vulnerabilities, and a team member manually analyzes and tries to exploit them. 
The 4-1 Practice Security Guidelines (SG) demands documentation  which is inherent a manual task. Guidelines describes how to securely integrate, configure and maintain the product. For large systems, such documents are of extreme value.

Note that continuous delivery represent automation until the release stage, while continuous deployment until Production. 
\vspace{-5pt}
\subsubsection{Operate and Monitor:} The product is available for Customer use.  Security monitoring, security testing and compliance checks are highly automated. Monitoring activities are part of Maintenance and artifacts belong to the Analytics repository. 

Security-related issues are tracked and redirected to the Concept stage where the countermeasure is packed into an Update. This is known as “Continuous Improvement and Feedback”. The 4-1 standard practice Security Update Management (SUM) demands synchronization of updates and roll-out tracking.

\section{Evaluation}\label{sec:evaluation}
We performed a qualitative study at Siemens AG, a large company that contributes to the IEC 62443-4-1, and where DevOps practices are applied to develop industrial systems. Based on current practices for empirical studies \cite{Wagner2019ChallengesIS,shull:2007:guide}, we interviewed practitioners who are part of projects related to security compliance and DevOps. 
The objective of the study is to answer the following research questions:

\begin{itemize}
    \item \textbf{RQ1.} How do practitioners perceive the precision of the automation classification criteria and automation capabilities of the 4-1 standard?
    \item \textbf{RQ2.} How do practitioners perceive the usefulness of our approach artefacts, specifically the (S\textsuperscript{2}C) DevOps Pipeline for the 4-1 security standard?
    \end{itemize}
  
\subsection{Study Design}

\subsubsection{Subjects}
In this study, we target practitioners that are knowledgeable of security compliance issues in DevOps projects, and whose products must be compliant with the 4-1 standard. As DevOps is recently being adopted for ICS, we conducted the study in a large industrial company. The subject group involves 7 participants developing products for industries like digital manufacturing, smart infrastructure, or healthcare. The number of subjects reflects the growing status of security compliance DevOps for ICS. In addition, this number is aligned with the current state-of-practice for comparable empirical studies on the field of security with restricted environments (containing similar sampling sizes between 5 and 11 practitioners (c.f. \cite{ben2017empirical})).
 
Participants are experienced professionals, with different levels of knowledge and expertise in the IEC 62443-4-1 standard and DevOps. For anonymity reasons, we excluded company roles since the number of experts in the field is limited. Table \ref{tab:participants} shows how the interviewees' perceived their level of expertise. None of the participants has classified himself as an expert in the given topic. One participant described himself as beginner in security tools, considering, their concentration is processes and people aspects, rather than the technology aspect of security. 

\begin{table}[h!]
\centering
\caption{Interviewees Perception of their Knowledge Level in study related topics}
\begin{tabular}{ccccc}
\hline
- & Beginner & Medium & Advanced & Expert \\ \hline
4-1 Standard & 2 & 3 & 2 & 0 \\ \hline
DevOps & 2 & 2 & 3 & 0 \\ \hline
DevSecOps & 2 & 2 & 3 & 0 \\ \hline
Security tools & 1 & 3 & 3 & 0 \\ \hline
\end{tabular}
\label{tab:participants}
\end{table}

\subsubsection{Interviews}
The interviews involved one participant and two interviewers. A interview protocol was prepared and is available in the online material. As preparation, specific parts of the 4-1 standard are selected to keep the time of the interviews short. Subjects received handouts of the selected parts that correspond to the artefacts presented in \cref{fig:03_integration}. The interviews took an average of one hour, the flow is the following:

\textbf{Introduction} (5 to 10 minutes) 
    Subjects are informed about the goal and the flow of the interview. We introduce the integration approach, the protocol rigor and measures to avoid bias.

\textbf{Part 1: Automation Criteria} (10 to 15 minutes)
    Aided by handouts, subjects provide opinions on the clarity of the categories in the automation criteria, namely: human task, transparency, partial automation, tools possible and complete automation.

\textbf{Part 2: 4-1 Automation Capabilities in detail} (20 to 30 minutes) Subjects are introduced to parts of the 4-1 pipeline specifications for specific selected 4-1 practices. Handouts have sets of activities that are highlighted. Interviewees provide opinions on the precision of the automation level for the activities and (when applicable) selected tools.  They are asked to improve the list of tools. 

\textbf{Part 3: 4-1 Automation Capabilities global and compliant pipeline} (10 to 20 minutes) 
    Subjects take a brief introduction on the global and per practice automation statistics of the 4-1 standard (see \cref{fig:automation_stats_practice}). They provide opinions of the expectations of automation capabilities. Later, they visualize the  S\textsuperscript{2}C DevOps Pipeline for the 4-1 standard  (\cref{fig:s2c_pipeline}) and discuss the usability, benefits, and drawbacks. 
    
\subsection{Results}

\subsubsection{Automation Criteria} The majority of the interviewees (6 of 7) found the automation criteria to be precise, however, some participants argued about the meaning of "\textit{Complete}" automation. They argue that any tool would require some human effort to configure it. 
Some quotes that resulted from the interviews are listed in Table \ref{tab:participant_quotes_automation}.  

\begin{table}[!htbp]
	\centering
	\caption{Interview Quotes regarding Automation Criteria}
	\scalebox{0.8}{	\begin{tabular}{C{2cm}L{16cm}}
			\toprule
\textbf{Interview No.} & \textbf{Interviewee Quote} \\ \midrule

2 & You might come into discussion with technical practitioners about the meaning of automation. \\ \midrule

5 & For me "Partial" and "Transparency" are the same classification; You will always have visualization and it's not relevant for automation. \\ \midrule

6 & What is the meaning of complete automation? \\ \midrule
7 & The automation criteria would be sufficient to capture the needed information.\\ \midrule
\bottomrule

\end{tabular}}
	\label{tab:participant_quotes_automation}
\end{table}

\subsubsection{Automation Capabilities in detail} The 4-1 standard pipeline specifications seem to be applicable as the participants generally agreed on the automation level, as well as some the selected tools. This is influenced by subjects' levels of awareness of the selected tools to fulfil the 4-1 standard activities. All subjects were aware about the mentioned security tools and their applicability for continuous integration and delivery pipelines. We collected very few extra tools, but received feedback on the applicability of same tools for other use cases, and reference to security tools expert to validate the tools list.

An advanced practitioner suggested to provide catalogs that enable easy discovering of new or alternative tools per activity.

\subsubsection{Automation capabilities global and compliant pipeline} Our interviewees would have expected less activities in the category \textit{Tool possible} and more in \textit{Complete} automation. However, they pointed out that the results seem realistic. Nevertheless, all subjects showed surprise with the automation capabilities for the SM practice. Management is more a human-related practice. Later, after a hint, interviewees with medium and advanced 4-1 knowledge remembered that the 4-1 SM Practice contains third party components security verification which is highly automatable and several tools are available.  About the (S\textsuperscript{2}C DevOps Pipeline, they are asked if it would be helpful for building compliant pipelines. Our interviewees were divided between its suitability for building or evaluating pipelines. Selected quotes are listed in \cref{tab:participant_quotes_pipeline}.


\begin{table}[!htbp]
	\centering
	\caption{Interview Quotes regarding the S\textsuperscript{2}C DevOps Pipeline}
	\scalebox{0.8}{	\begin{tabular}{C{2cm}L{16cm}}
			\toprule
\textbf{Interview No.} 
& \textbf{Interviewee Quote}
\\ \midrule

3 & You could see where you can use a tool for automation. \\ \midrule

5 & A practitioner could discover new tools and easily spot a replacement for the same task. \\ \midrule

6 & You'd have to try it out when you build a task. \\ \midrule

6 & Ignoring the 4-1 standard, it would be a good source of tools to fulfil the tasks; Finding the right tool for the right task. \\ \midrule

7 & Could be helpful to evaluate a pipeline. \\
\bottomrule
\end{tabular}}
	\label{tab:participant_quotes_pipeline}
\end{table}

\subsection{Threats to Validity}
\noindent
To support the representativeness, we confirmed the participants' background and suitability to answer our suggestions at the beginning of the interviews. The sampling size of 5 to 11 practitioners is further in tune with current security research and comparable studies that have a restricted environment \cite{ben2017empirical}.

To further mitigate that participants alter their answers, we informed them about how the anonymity of their answers was preserved. Additionally, the interviewers explained the protocol and took measures to not influence answers like remaining neutral and in silence. 

Finally, to avoid false interpretations and overlook information, two interviewers participated, one applying the protocol and other one controlling rigor. Later, the results were validated by two different reviewers.

\section{Discussion} \label{sec:discussion}
\subsection{Summary of conclusions}

Security standards are described as linear processes, thus, are not prepared for DevOps iterative processes like continuous integration, delivery, or deployment. Our work translates security standard requirements into a pipeline-ready language describing: activity, input, output, automation capability, and when available, the list of security tools (see examples SI-t5 and SI-t6 in \cref{fig:excerpt_specification}). 

Security compliance activities are perceived as overload for DevOps teams, mostly including documentation tasks that are not perceived as customer value and that increase the lead time. Indeed, in industrial environments adopting agile and DevOps practices, practitioners constantly argue that security compliance reduces agility. Our work reveals that 31\% of the 4-1 standard security activities can be automated into a pipeline. In continuous software engineering, achieving compliance for this third can be an initial objective to improve the product development process. 

We also identified 38\% activities that can not be automated. Including them in DevOps implies better collaboration between security experts and DevOps teams. We recommend to improve security skills in the teams in order to avoid bottlenecks due to lacks of security experts. These human tasks may be heavy for the backlog and include documentation, e.g. writing security guidelines or drawing security architecture diagrams. Although, they are not typical for agile approaches, they are still highly demanded by standards. 

\subsection{Limitations}
Limitations to our approach arise from our focus on a ICS specific standard. However, since the 4-1 is derived from the ISO27034 - the main standard for secure software development - our approach applies to other domains as well. Also, it can be used to introduce security standards that demand process requirements for product delivery. 
Main limitations of using the 4-1 standard are: a) it does not cover all aspects of secure operations e.g. secure hardening, b) the 4-1 was released in 2018, therefore knowledgeable practitioners are very rare and belong to industrial companies who can be certified but not necessarily based on DevOps practices. This also forces evaluation endeavours to be performed in restricted environments and to remain qualitative of nature. 

\subsection{Impact} 

\subsubsection{For industry professionals}
DevOps practitioners can use our artefacts to establish continuous security improvement roadmaps, e.g. the first iteration could be to introduce in their DevOps pipelines the \textit{complete} automated security activities, subsequently, the \textit{partial automation} activities, and so on.
Currently, in the industrial company, S\textsuperscript{2}C DevOps Pipeline is applied as common language to bridge communication between security experts, dev and ops teams. We forsee the applicability of this pipeline specification as a pre-set documentation template for their pipelines.
Further, compliance auditors can evolve our artefacts as templates to perform compliance assessments of DevOps pipelines with regard to the 4-1 standard.
 
Security experts can use the artefacts to provide certain recommendations on how to apply security activities in a more automated way, while avoiding security tasks not fitting with DevOps engineering practices. 

Moreover, safety practitioners may replicate the approach for functional safety standards like the ISO26262 or IEC61508 \cite{iso26262,iec61508} .

\subsubsection{For researchers}
We extend the work on DevOps security for regulated environments with a systematic method and automation criteria to evaluate how compliance activities can influence DevOps lead time and team collaboration. Using our artefacts, researchers can discover directions for the field like: a) applying machine learning to build up tools in the category "Tool possible", e.g. to recommend treatments to security issues, or b) verifying if activities are clear enough to design and build tools.

\subsubsection{For standardization organizations}
Artefacts serve as recommended templates to document compliance and achieve the same level of understanding among practitioners of different fields. Our approach is one way to map and obtain evidence for ambiguous statements in standards. Besides, the industrial scenario may serve as a shareable case study to start discussions on how to achieve the same level of security for several ICS. At the end, this is the aim of standards: to pursue large application of security best practices so that ICS in global critical infrastructures remain resilient to attacks.

\section{Conclusion} \label{sec:conclusion}
The present work demonstrates the potential of systematic integration of standard requirements into DevOps pipelines. Considering that pipelines are based on automation, our work analyzes also automation capabilities of security standard requirements as a key element to support DevOps aims. The analysis describes which requirements of the standard can be automated, either fully or partially. We provide an overview of the results and identified gaps. These gaps are  requirements that can not be automated and have to be performed manually.

The qualitative evaluation in a large industrial company shows evidence that a S\textsuperscript{2}C DevOps pipeline specification is useful to assess security compliance of DevOps pipelines. Moreover, the automation capabilities of the standard requirements serves to build security compliant pipelines with a potential of at least partially  automating over 60\%  activities arising from standard requirements. 

Our contributions to the field of security compliant DevOps are as follows:
\begin{enumerate}
  \item We developed an approach to integrate security standards into pipelines and we validated it through an instance for the IEC~62443-4-1 standard that regulates the ICS domain. The main artefact of this approach is the Security Standard Compliant S\textsuperscript{2}C DevOps Pipeline Specification. It describes how the practices of the security standard fit into the DevOps Pipeline.
  \item We analysed and documented the automation capabilities of the 4-1 standard, in the context of a large industrial company that operates in the ICS market. We found that the automation extent of this standard is 31\% \textit{Complete}, it means that 31\% of the 4-1 requirements can be fully automated.  38\% of the 4-1 requirements are manual tasks that have to be executed by a human expert. The remainder has the potential to be at least partially automated with future tools and techniques. 
\end{enumerate}

Currently, we use the S\textsuperscript{2}C DevOps pipeline specification (\cref{fig:s2c_pipeline}) to raise awareness for the compliance flow in DevOps as well as possibilities to achieve compliance automation. Also, we experiment with building up automation pipelines for security compliance aided by the 4-1 standard pipeline specifications. Moreover, we look into building tools to increase the percentage of 4-1 requirements that can be fully automated
Finally, given our positive experience with the case at hands, we cordially invite researchers and practitioners in joining our endeavour to further scaling up our work to other standards and domains.

\section{Acknowledgements}  
This work is partially funded by portuguese national funds through FCT - Fundação para a Ciência e Tecnologia, I.P., under the project FCT UIDB/04466/2020. Furthermore, the third author thanks the Instituto Universitário de Lisboa and ISTAR-IUL, for their support.

%
%
%

\vspace{30pt}
\begingroup
\let\clearpage\relax
\bibliographystyle{splncs04}
\bibliography{bibliography.bib}

\begin{thebibliography}{10}
\providecommand{\url}[1]{\texttt{#1}}
\providecommand{\urlprefix}{URL }
\providecommand{\doi}[1]{https://doi.org/#1}

\bibitem{allspaw:2009:ten}
Allspaw, J., Hammond, P.: 10+ deploys per day: Dev and ops cooperation at
  flickr. In: Velocity: Web Performance and Operations Conference. O'Reilly
  (2009)

\bibitem{beck2001manifesto}
Beck, K., Beedle, M., Van~Bennekum, A., Cockburn, A., Cunningham, W., Fowler,
  M., Grenning, J., Highsmith, J., Hunt, A., Jeffries, R., et~al.: Manifesto
  for agile software development  (2001)

\bibitem{ben2017empirical}
Ben~Othmane, L., Jaatun, M.G., Weippl, E.: Empirical Research for Software
  Security: Foundations and Experience. CRC Press (2017)

\bibitem{bird2016devopssec}
Bird, J.: Security as Code: Security Tools and Practices in Continuous
  Delivery, chap.~4, pp. 32--36. O'Reilly Media, Incorporated (2016)

\bibitem{dora_report}
DORA: Accelerate: State of devops (2019),
  \url{https://services.google.com/fh/files/misc/state-of-devops-2019.pdf}

\bibitem{gartner_report}
Gartner: 10 things to get right for successful devsecops (2017),
  \url{https://www.gartner.com/en/documents/3811369/10-things-to-get-right-for-successful-devsecops}

\bibitem{hsu2018hands}
Hsu, T.H.C.: Hands-On Security in DevOps: Ensure continuous security,
  deployment, and delivery with DevSecOps. Packt Publishing Ltd (2018)

\bibitem{humble2010continuous}
Humble, J., Farley, D.: Continuous Delivery: Reliable Software Releases through
  Build, Test, and Deployment Automation. Pearson Education (2010)

\bibitem{iec61508}
IEC: 61508 - functional safety. International Electrotechnical Commission
  (2010)

\bibitem{iec4_1}
(IEC), I.E.C.: 62443-4-1. {Security for industrial automation and control
  systems Part 4-1 Product security development life-cycle requirements} (2018)

\bibitem{iso26262}
ISO: 26262 - road vehicles — functional safety. International Standards
  Organization  (2011)

\bibitem{jaatun:2017:devops}
Jaatun, M.G., Cruzes, D.S., Luna, J.: Devops for better software security in
  the cloud invited paper. In: Proceedings of the 12th ARES. ACM, New York, NY,
  USA (2017)

\bibitem{Jabbari:2016}
Jabbari, R., bin Ali, N., Petersen, K., Tanveer, B.: What is devops?: A
  systematic mapping study on definitions and practices. In: Proceedings of
  Workshop XP. ACM, USA (2016)

\bibitem{kim:2018:phoenix}
Kim, G., Behr, K., Spafford, G.: The Phoenix Project: A Novel about IT, DevOps,
  and Helping Your Business Win. IT Revolution Press (2018)

\bibitem{kim:2016:devops}
Kim, G., Humble, J., Debois, P., Willis, J.: The DevOps Handbook:: How to
  Create World-Class Agility, Reliability, and Security in Technology
  Organizations. IT Revolution Press (2016)

\bibitem{laukkarinen:2018:regulated}
Laukkarinen, T., Kuusinen, K., Mikkonen, T.: Regulated software meets devops.
  Information and Software Technology  \textbf{97} (2018)

\bibitem{leite:2019:survey}
Leite, L., Rocha, C., Kon, F., Milojicic, D., Meirelles, P.: A survey of devops
  concepts and challenges. vol.~52. Association for Computing Machinery, New
  York, NY, USA (Nov 2019)

\bibitem{lwakatare2015dimensions}
Lwakatare, L.E., Kuvaja, P., Oivo, M.: Dimensions of devops. In: International
  conference on agile software development. pp. 212--217. Springer (2015)

\bibitem{michener:2016:mitigating}
{Michener}, J.R., {Clager}, A.T.: Mitigating an oxymoron: Compliance in a
  devops environments. In: 2016 IEEE 40th COMPSAC. vol.~1, pp. 396--398 (2016)

\bibitem{mohan2016secdevops}
Mohan, V., Othmane, L.B.: Secdevops: Is it a marketing buzzword?-mapping
  research on security in devops. In: 11th ARES. pp. 542--547. IEEE (2016)

\bibitem{morales:2020:guide}
Morales, J., Turner, R., Miller, S., Capell, P., Place, P., Shepard, D.: Guide
  to implementing devsecops for a system of systems in highly regulated
  environments. Tech. Rep. CMU/SEI-2020-TR-002, SEI, Carnegie Mellon
  University, Pittsburgh, PA (2020)

\bibitem{Moyon:2018}
Moy{\'o}n, F., Beckers, K., Klepper, S., Lachberger, P., Bruegge, B.: Towards
  continuous security compliance in agile software development at scale. In:
  Proc. of RCoSE. ACM (2018)

\bibitem{poppendieck:2003:lean}
Poppendieck, M., Poppendieck, T.: Lean Software Development: An Agile Toolkit.
  Agile Software Development Series, Pearson Education (2003)

\bibitem{sansposter}
SANS: Sans secure devops tooolchain and securing web application technologies
  checklist (2018)

\bibitem{shahin2017continuous}
Shahin, M., Babar, M.A., Zhu, L.: Continuous integration, delivery and
  deployment: a systematic review on approaches, tools, challenges and
  practices. IEEE  (2017)

\bibitem{shull:2007:guide}
Shull, F., Singer, J., Sj{\o}berg, D.I.: Guide to advanced empirical sw
  engineering. Springer (2007)

\bibitem{sonatype_report}
Sonatype: Devsecops community survey, 2019 (2019)

\bibitem{rahman:2016:software}
Ur~Rahman, A.A., Williams, L.: Software security in devops: Synthesizing
  practitioners’ perceptions and practices. In: Proceedings of International
  Workshop CSED. ACM, USA (2016)

\bibitem{Wagner2019ChallengesIS}
Wagner, S., Fern{\'a}ndez, D.M., Felderer, M., Graziotin, D., Kalinowski, M.:
  Challenges in survey research. ArXiv  \textbf{abs/1908.05899} (2019)

\bibitem{yasar:2017:implementing}
Yasar, H.: Implementing secure devops assessment for highly regulated
  environments. In: Proceedings of the 12th ARES. ACM, USA (2017)

\bibitem{yasar:2016:where}
Yasar, H., Kontostathis, K.: Where to integrate security practices on devops
  platform. Int. J. Secur. Softw. Eng.  \textbf{7}(4) (Oct 2016)

\end{thebibliography}
\endgroup
\end{document}